\documentclass[prb,onecolumn,showpacs,superscriptaddress]{revtex4-1}
\usepackage{amsmath}
\usepackage{graphicx}

\newcommand{\GL}{\mathrm{\scriptscriptstyle GL}}
\newcommand{\Gs}{{\scriptscriptstyle G}}

\newcommand{\DOS}{\mathrm{\scriptscriptstyle DOS}}
\newcommand{\MT}{\mathrm{\scriptscriptstyle MT}}
\newcommand{\AL}{\mathrm{\scriptscriptstyle AL}}
\newcommand{\cp}{\mathrm{ cp}}

\newcommand{\bnabla}{{\boldsymbol{\nabla}}} \newcommand{\Tr}{\mathrm{Tr}} \newcommand{\Dk}{\check{\Delta}_{\cal K}} \newcommand{\Qk}{\check{Q}_{\cal K}} 
 \newcommand{\Ak}{\check{\mathbf{A}}_{\cal K}} \newcommand{\Si}{\check{\Xi}}
\newcommand{\cK}{\cal K}

\newcommand{\bx}{{\mathbf x}}
\newcommand{\br}{{\mathbf r}}

\newcommand{\bq}{{\mathbf q}}

\newcommand{\dif}{{\mathrm d}}

\def\Xint#1{\mathchoice
{\XXint\displaystyle\textstyle{#1}}%
{\XXint\textstyle\scriptstyle{#1}}%
{\XXint\scriptstyle\scriptscriptstyle{#1}}%
{\XXint\scriptscriptstyle\scriptscriptstyle{#1}}%
\!\int}
\def\XXint#1#2#3{{\setbox0=\hbox{$#1{#2#3}{\int}$}
\vcenter{\hbox{$#2#3$}}\kern-.5\wd0}}

\def\dashint{\Xint-}

\newcount\minute
\newcount\hour
\newcount\hourMins
\def\now
{
  \minute=\time
  \hour=\time \divide \hour by 60
  \hourMins=\hour \multiply\hourMins by 60
  \advance\minute by -\hourMins
  \zeroPadTwo{\the\hour}:\zeroPadTwo{\the\minute}
}\def\timestamp
{
  \today\ \now
}
\def\today
{
  \zeroPadTwo{\the\day}-\zeroPadTwo{\the\month}-\the\year%
}
\def\zeroPadTwo#1%
{
  \ifnum #1<10 0\fi
  #1%
}

\date{\timestamp}

\begin{document}

\title{Fluctuation-induced noise in out-of-equilibrium disordered superconducting films}
\author{Aleksandra Petkovi\'{c}}\email{Corresponding author. E-mail address: alpetkovic@gmail.com}
\affiliation{Laboratoire de Physique Th\'{e}orique, IRSAMC, CNRS and Universit\'{e} de Toulouse, UPS, F-31062 Toulouse, France}
\affiliation{Laboratoire de Physique Th\'{e}orique et Hautes Energies, Universit\'{e} Pierre et Marie Curie and CNRS UMR 7589, 4 place Jussieu, 75005 Paris, France
}
\affiliation{Laboratoire de Physique Th\'{e}orique-CNRS, Ecole Normale Sup\'{e}rieure, 24 rue Lhomond, 75005 Paris, France}

\author{Valerii M. Vinokur}
\affiliation{Materials Science Division, Argonne National Laboratory, Argonne, Illinois 60439, USA}

\begin{abstract}
We study out-of-equilibrium transport in disordered superconductors close to the superconducting transition. We consider a thin film connected by resistive tunnel interfaces to thermal reservoirs having different chemical potentials and temperatures. The nonequilibrium longitudinal current-current correlation function is calculated within the nonlinear sigma model description and nonlinear dependence on temperatures and chemical potentials is obtained. Different contributions are calculated, originating from the fluctuation-induced suppression of the quasiparticle density of states, Maki-Thompson and Aslamazov-Larkin processes.  As a special case of our results, close-to-equilibrium we obtain the longitudinal ac conductivity using the fluctuation-dissipation theorem.
\end{abstract}



\pacs{74.40.-n, 74.25.F-}
\maketitle

\section{Introduction}

An equilibrium in nature is rare and it is rather an exception than the rule. A vast majority of natural processes, ranging from large-scale flows in the atmosphere to the electric charge transfer found throughout the technological realm, are out-of-equilibrium processes.
Yet, the physics of the equilibrium state is much more studied and better understood. A major difficulty impeding the similar progress in the research of the nonequilibrium processes is that while all thermodynamics-based science rests on the law that, in equilibrium, any system assumes the state with the minimal free energy, the equally powerful and fundamental principles that govern the far-from-equilibrium behaviors still wait to be revealed. There has been an important advance both theoretical and experimental\cite{Blanter,Nazarov,Kogan,Jong+,Kopnin} employing a variety of approaches to out-of-equilibrium problems in
electronic  systems\cite{Kopnin,Khlus,Lesovik,Buttiker,Beenakker,Shulman,Nagaev,NagaevSc,Altshuler,Gefen};
among them, those based on the Keldysh technique\cite{Keldysh,L_Kamenev}
seem to appear the most promising as paving a way towards general method which would allow
treating interacting nonequilibrium systems on a common fundamental ground.

In this work we undertake the study of a disordered superconducting system employing the theory that has been proven to effectively tackle the low-energy excitation physics, the Keldysh nonlinear sigma model \cite{Feigelman+00}.
More specifically, we focus on the nonequilibrium phase transition between the normal and the superconducting state.
In a vicinity of the transition, when the system is in the normal state, the behavior of the system is governed by fluctuations of the superconducting order parameter. Fluctuation-induced short-living Cooper pairs are formed and contribute critically both into the thermodynamic and transport characteristics of the systems\cite{Varlamovbook}. In disordered thin films the temperature range where fluctuations are essential is determined
by the sheet resistance (being in any case significantly larger than the one of bulk superconductors\cite{Aslamazov1,Aslamazov2,Maki,Varlamovbook}) and depends on the
particular process involved, extending often to temperatures well above the superconducting transition
temperature\cite{Varlamovbook,Baturina2012}.

Here we study the influence of superconducting fluctuations on dynamic properties of a thin film in the
fluctuational region of a normal state. The film is driven out of the equilibrium due to contacts with thermal reservoirs having different temperatures and chemical potentials.
The Keldysh Ginzburg-Landau-like action under nonequilibrium conditions and several effects of nonequilibrium superconducting fluctuations  were addressed in Refs.~\onlinecite{NikolayEPL,PRL,PRBmy} for this setting.  Out-of-equilibrium fluctuation contributions to the dc electrical conductivity
were calculated in Ref.~\onlinecite{PRBmy}.
Importantly, under nonequilibrium conditions the fluctuation-dissipation theorem (FDT) is generally
violated and there is no fundamental relation between the current fluctuations (i.e.~noise) and conductivity,
in contrast to close-to-equilibrium conditions where this relation holds.
Therefore, unlike in equilibrium systems, the nonequilibrium noise is not fully tied to the conductivity and can carry additional information, not contained in the conductivity. This poses a problem of the independent calculation of noise in out-of-equilibrium state.

Shot noise in noninteracting diffusive electron system was studied in
Refs.~\onlinecite{Nagaev,Altshuler}.
The influence of the Coulomb interaction on shot noise in disordered systems was analyzed in
Refs.~\onlinecite{Beenakker,Nagaev,Kozub,Gefen}, while the influence of  Bardeen-Cooper-Schrieffer (BCS)
interaction on shot noise due to electric current flow above the critical temperature  was considered only up to
the second order in the electric field in Ref.~\onlinecite{NagaevSc}.
In the present work, we focus on a film above the nonequilibrium superconducting transition and
calculate {nonlinear} dependence of the Nyquist noise on the temperatures and chemical potentials
of the thermal baths that are in contact with the film, see Fig.~\ref{fig:film}.

The paper is organized as follows.  In Sec.~\ref{sec:NLSM} we
introduce the nonlinear sigma model for superconductors within the Keldysh technique. In Sec.~\ref{sec:noninteracting} we find the current correlation function for the case of noninteracting electrons in a nonequilibrium disordered thin film. We further consider the BCS interaction. In Sec.~\ref{sec:fluctuations1} we introduce nonequilibrium
fluctuation propagators. Then we proceed with calculations of different contributions to the current-current correlation function
caused by superconducting fluctuations: the density of states contribution is calculated in Sec.~\ref{sec:DOS}, the Maki-Thompson in Sec.~\ref{sec:MT} and the Aslamazov-Larkin in Sec.~\ref{sec:AL}.
In Sec.~\ref{sec:conclusions} we summarize our results.
Some calculation details are given in Appendices.

\section{Keldysh formalism: disordered superconductors\label{sec:NLSM}}

In this section we introduce the notation and provide the basic equations  needed for the calculation of the current-current correlation function. We present the model, discuss its applicability and explain the procedure that allows us to analyze the fluctuations of the superconducting order parameter in the metallic state. In order to treat the nonequilibrium physics, we employ the Keldysh technique \cite{Keldysh}. We start with the nonlinear sigma model and then further discuss the calculation of the current-current correlations.

\subsection{Nonlinear sigma model }

The nonlinear sigma model can be used to describe the low-energy physics for superconductors with BCS interaction in the presence of short-range quenched disorder (see Appendix \ref{app:model}).  The partition function takes the form $Z=\int \mathcal{D}Q\;\mathcal{D}\Delta\exp\{i S[\Qk,\Dk]\}$, where the action $S$ consists of three parts\cite{Feigelman+00,KamenevAndrev}
\begin{align}
\label{eq:S}
&S[\Qk,\Dk]=S_{\Delta}+S_{\phi}+S_{Q}.
\end{align}
Here and in the following we set $\hbar=c=k_B=1$. The fields $\Delta$ and $Q$ are Hubbard-Stratonovich fields introduced to decouple the four-fermion terms originating from the Hamiltonian describing the BCS interaction and disorder, respectively\cite{Feigelman+00,KamenevAndrev}.
The contributions to the action (\ref{eq:S}) are given by
\begin{align}
S_{\Delta}=&-\frac{\nu}{2\lambda}\Tr[\Dk^{\dagger}\check{Y}\Dk],\quad S_{\phi}=\frac{e^2\nu}{2}\Tr[\check \phi_{\mathcal{K}}\check{Y}\check \phi_{\mathcal{K}}],\\
\label{q:S_Q}
S_{Q}=&\frac{i \pi \nu}{4}\Tr[D (\partial_{\bf{r}}\Qk)^2-4 \Si\partial_t\Qk
-4ie\check{\phi}_{\mathcal{K}}\Qk+4i\Dk\Qk ].
\end{align}
Here $D$ is the diffusion coefficient, and it carries information about the disorder. The bare single particle density of states at the Fermi level per one spin projection is denoted by $\nu$. The superconductive coupling constant $\lambda$ is positive. The matrix field $\check{Q}$ satisfies the nonlinear relation $\check Q^2=1$. The check symbol $\check\ $ denotes  $4\times4$ matrices that are defined in the tensor product of the Keldysh and Nambu spaces. The former and the latter are spanned by the Pauli matrices $\hat{\sigma}_i$ and $\hat{\tau}_i$, $i\in\{0,x,y,z\}$, respectively, and we define $\check{Y}=\hat{\sigma}_x\otimes \hat{\tau}_0$, $\Si=\hat{\sigma}_0\otimes\hat{\tau}_z$. One uses different notation for the same matrices $\hat{\sigma}_i=\hat{\tau}_i$ for convenience, and $\hat{\sigma}_0=\mathrm{diag}(1,1)$. We assume implicitly the multiplication in the time-space, and ``Tr'' includes the integration over the real space. The subscript $\cK$ denotes the gauge transformed fields $\check\phi_{\cK}=\check{\phi}-\partial_{t}\check{\cK}$, $ {\Ak}={\check{\mathbf{A}}} +{\bf{\bnabla}}{\check{\cK}}$ and $\check{\cK}=\left(k^{cl}\hat{\sigma}_{0}+k^{q}\hat{\sigma}_{x}\right)\otimes\hat{\tau}_{0}$.
The fields $\check A$ and  $\check \phi$ are defined in same way as $\check{\cK}$, where $A$ and $\phi$ are the vector and the scalar potential, respectively. The field $\check{\Delta}$ is given by
$\check{\Delta}= \left(\Delta^{cl}\hat{\sigma}_{0}+ \Delta^{q}\hat{\sigma}_{x}\right)\otimes\hat{\tau}_{+}-\mathrm{H.c.}$, and $\Dk(\br,t)=\exp{\left[ie \Si{\check{\cK}}(\br,t)\right]}\check{\Delta}\exp{\left[-ie\Si{\check{\cK}}(\br,t)\right]}$.
$\Qk$ is defined in the same way. We have also defined $\hat{\tau}_{\pm}=(\hat{\tau}_x\pm i\hat{\tau}_y)/2$.
The quantum (q) and classical (cl) components of the fields are respectively defined as the half-sum and the half-difference of the field values at the lower and the upper branches of the Keldysh time-contour. The field $\Delta^{cl}$ becomes the superconducting order parameter at the mean-field (saddle-point) level, while the saddle point equation for $\check Q$ produces the Usadel quasiclassical equations, where $\check Q$ plays the role of the quasiclassical Greens function. The covariant spatial derivative is given by $
\partial_{\mathbf{r}}\check{Q}_{\cK}= \bnabla_{\mathbf{r}}\check{Q}_{\cK}- ie[\check{\Xi}\check{\mathbf{A}}_{\cK},\check{Q}_{\cK}]$.
We stress that the nonlinear sigma model action $S$ captures the low-energy physics at energy scales much smaller that the elastic scattering rate. It is valid in the limit when the lifetime of fluctuating Cooper pairs is much greater than the elastic scattering time.

We further explain the strategy to treat the superconducting fluctuations. First we find the saddle point equation for $\check{Q}$ of the action (\ref{eq:S}), in the absence of the BCS interaction (i.e.~$\lambda=0$). It reads as \cite{first,second,KamenevAndrev, Feigelman+00}
\begin{align}\label{eq:saddle}
\check \Lambda&=\check{\mathcal{U}}\check{\Lambda}_0 \check{\mathcal{U}}^{-1},\quad\quad\check{\Lambda}_0=\hat{\sigma}_z\otimes\hat{\tau}_z,\\
\label{lambda}
\check{\mathcal{U}}_{t,t'}(\br)&=\check{\mathcal{U}}_{t,t'}^{-1}(\br)=
\left(\begin{array}{cc}\delta(t-t'-0)\hat{\tau}_0&
\hat{F}_{t,t'}(\br)\\
0&-\delta(t-t'+0)\hat{\tau}_0\end{array}\right),\\
\hat{F}_{t,t'}(\br)&=\begin{pmatrix}
F_{t,t'}^e(\br)& 0\\
0& F_{t,t'}^h(\br)
\end{pmatrix}.
\end{align}
After Wigner transforming ${F}_{t,t'}^{e/h}(\br)$ we obtain $F^{e/h}_{\epsilon}(\br,t)$ that can be related to the quasiparticle electron/hole distribution functions $f^{e/h}_{\epsilon}(\br,t)$ via $F^{e/h}_{\epsilon}(\br,t)= 1-2f^{e/h}_{\epsilon}(\br,t)$. One then considers massless fluctuations around the normal-metal saddle point solution, since massive modes can be integrated out in the Gaussian approximation and lead to unimportant renormalization of the parameters in the action. The massless fluctuations satisfy $\check Q^2=1$ and are conveniently parameterized as \cite{Feigelman+00}
\begin{align}\label{parametrization}
\check{Q}_{\cK}(\mathbf{r}) &= e^{- \check{W}(\mathbf{r})/2}\, \check{\Lambda}(\br)\,\, e^{\check{W}(\mathbf{r})/2 },\quad \check{W}=\check{\mathcal{U}}\check{\mathcal{W}} \check{\mathcal{U}}^{-1},\\
\label{eq:W}
\check{\mathcal{W}}&=\left(\begin{array}{cc}
w\tau_{+} - w^{*}\tau_{-} & w_{0}\tau_{0}+w_{z}\tau_{z}\\
\bar{w}_{0}\tau_{0}+\bar{w}_{z}\tau_{z} & \bar{w}\tau_{+} -
\bar{w}^{*}\tau_{-}
\end{array}\right),
\end{align}
such that $\check{W}\check{\Lambda}+\check{\Lambda}\check{W}=0$. Here we introduced four real fields $w^{\alpha}_{tt'}(\mathbf{r}),\bar{w}^{\alpha}_{tt'}(\mathbf{r})$
with $\alpha=0,z$ representing diffuson degrees of freedom and the two complex
fields $w_{tt'}(\mathbf{r}),\bar{w}_{tt'}(\mathbf{r})$ for Cooperon degrees of freedom. One now substitutes $\check{Q}$ matrix given by Eq.(\ref{parametrization}) into the action (\ref{eq:S}) and requires that the terms linear in $\check{W}$ vanish. This leads to a kinetic equation that electron and hole distribution function have to satisfy. Next we switch on the BCS interaction assuming the system is in the normal state. In that case the average value of the superconducting order parameter is zero, and therefore one again obtains the normal-metal saddle point solution (\ref{eq:saddle}). In order to study the influence of the BCS interaction, one has to consider the fluctuations. Assuming that the system is not too  close to the transition, we take into account quadratic fluctuations around the metallic saddle point solution. Integrating them out, one obtains the effective action depending only on the superconducting order parameter and electromagnetic fields. This action allows us to treat the superconducting fluctuations in the normal metallic state. \cite{AlexPRB,PRBmy}

\subsection{Current-current correlation function\label{sec:current-current}}

Our aim in the following is to calculate the symmetrized two-operator current correlation function
\begin{align}
S(\br,t;\br',t')= \frac{1}{2}\langle {\mathbf{j}}_x(\br,t){\mathbf{j}}_x(\br',t')
+{\mathbf{j}}_x(\br',t'){\mathbf{j}}_x(\br,t)\rangle
\end{align}
under nonequilibrium conditions. It can be obtained differentiating the  nonlinear sigma model partition function
\begin{align}\label{eq:noisedef}
S(\br,t;\br',t')=-\frac{1}{4}\frac{\partial^2 Z}{\partial {\bf{A}}^q_x(\br,t)\partial{\bf{A}}^q_x(\br',t')}\big|_{{\bf{A}}^q=0,{\bf{A}}^{cl}=0}.
\end{align}
\begin{figure}
\includegraphics[width=0.35\columnwidth]{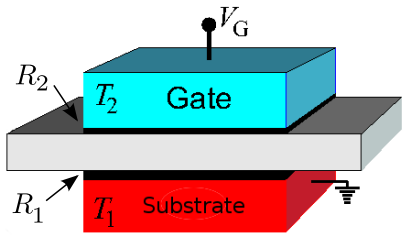}
\caption{Thin superconducting film is connected by tunneling interfaces to the gate and the substrate. The interfaces are characterized by the resistances $R_1$ and $R_2$. The substrate temperature is $T_1$, the gate temperature is $T_2$ and the gate voltage is $V_{G}$. The parameters $T_{1}$, $T_{2}$, $V_{G}$, and the resistances of the tunneling interfaces determine the quasiparticle distribution in the film and allow us to change it in a controlled way. \label{fig:film}}
\end{figure}
In the rest of the paper we focus on the particular system shown in Fig.~\ref{fig:film}. We consider a thin superconducting film connected by resistive interfaces to thermal reservoirs having different chemical potentials and temperatures. We assume that the system is in zero magnetic field and in the normal state but close to the transition into the superconducting state. We study a stationary situation  ($F^{e/h}_{t,t'}=F^{e/h}_{t-t'}$). Therefore, the noise depends only on the time difference $t-t'$. In the following we use the notation $F_{\epsilon}^{e/h}\equiv F^{e/h}(\epsilon)$ for the energy dependence of the distribution function.

There are four different contributions to the current correlation function
\begin{align}\label{eq:Sdef}
S(\br-\br',t-t')=&S_0(\br-\br',t-t')+S_{\DOS}(\br-\br',t-t')
+S_{\MT}(\br-\br',t-t')+
S_{\AL}(\br-\br',t-t'),
\end{align}
where $S_0$ denotes the noise in the noninteracting case ($\lambda=0$), and the other terms are contributions induced by superconducting fluctuations.
The main processes are: i)coherent Andreev
reflections of quasiparticles on the local fluctuations of the superconducting order parameter resulting into the so-called Maki-Thompson (MT) contributions\cite{Maki,Thompson}; ii) formation of Cooper pairs and their involvement into charge transfer is described by the Aslamazov-Larkin  (AL) corrections\cite{Aslamazov1}, and iii) the suppression of the single-particle density of states (DOS) due to quasiparticle participation in Cooper-pairing \cite{Varlamovbook}. Randomness in the production and dissociation of fluctuating Cooper pairs, and in other closely related processes discussed above, gives the corresponding contributions to the current noise.  In the following sections we evaluate and analyze the individual contributions. We focus on the regime where the superconducting fluctuations can be treated perturbatively, i.e., the fluctuation contribution are small compared to the noise $S_0$ in the noninteracting case. This implies that the film is not too close to the superconducting transition and that the physics is dominated by Gaussian fluctuations. Then one finds $S_{\mathrm{fluct}}/S_{0}\sim G_i=(\nu D d_f)^{-1}\ll 1$. Here $G_i$ is the Ginzburg number, $d_f$ is the film thickness and $S_{\mathrm{fluct}}=S_{\DOS}+S_{\MT}+S_{\AL}$ denotes the sum of the contributions induced by fluctuations of the superconducting order parameter.

\section{Noninteracting electrons\label{sec:noninteracting}}

We consider noninteracting electrons first and evaluate Eq.~(\ref{eq:noisedef}) taking into account nonequilibrium conditions. After differentiating the partition function with respect to the vector potential field in Eq.~(\ref{eq:noisedef}), we substitute the field $\check{Q}_{\cK}$ with its saddle-point solution $\check{\Lambda}$ given by Eq.~(\ref{eq:saddle}). We do not take here into account the fluctuations around $\check{\Lambda}$, since they  are responsible for weak-localization effects. Since the main aim of this paper is to find the noise close to the superconducting transition, and since the weak-localization contribution is a non-singular function in the vicinity of the transition, we do not consider it here. Then, we obtain the longitudinal current-current correlation function to be\cite{Nagaev}
\begin{align}\label{eq:noisenoninteracting}
S_0(\br-\br',\epsilon)=&2 \sigma_D T_0(\epsilon)\delta(\br-\br'),\\
T_0(\epsilon)=&\frac{1}{4}\int \dif \Omega \Big\{1-\frac{1}{2}\big[F^e({\Omega})F^e({\Omega+\epsilon})+
F^h({\Omega})F^h({\Omega+\epsilon}) \big]\Big\}.
\end{align}
Here the Drude conductivity is $\sigma_D=2e^2\nu D$ and the integration is always from $-\infty$ to $+\infty$ if not specified differently. In equilibrium, the distribution function is $F^{e/h}(\epsilon)=\tanh\left(\epsilon/(2T)\right)$ and the expression (\ref{eq:noisenoninteracting}) simplifies to
\begin{align}\label{eq:soeq}
  S^{eq}_0(\br-\br',\epsilon)=\delta(\br-\br')\sigma_D \epsilon \coth{\left(\frac{\epsilon}{2T}\right)}.
\end{align}
This universal relation between the conductivity and the noise is known as the FDT.  The microscopic details about the disorder strength are hidden in the diffusion constant, i.e., in $\sigma_D$. We stress that the range of applicability of the nonlinear sigma model is $\epsilon\ll \tau^{-1}$, where $\tau$ is the elastic scattering time. This explains why in Eq.~(\ref{eq:soeq}) appears the frequency independent Drude conductivity $\sigma_D=2e^2\nu D$.

Next we analyze Eq.~(\ref{eq:noisenoninteracting}) in out-of-equilibrium conditions. Then, the FDT is generally violated. We assume that the film is thin such that the Thouless energy corresponding to diffusion across the film  $E_T^\perp=D/d_f^2$, well exceeds all the relevant energy scales. Here $d_f$ denotes the film thickness. The current across the interface separating the substrate and the film is
$I=\int\dif\epsilon\left[F^e(\epsilon)-F^e_S(\epsilon)-F^h(\epsilon)+F^h_S(\epsilon) \right]/(4 e R_1)$,
where $R_1$ is the tunneling resistance per unit area characterizing the interface and the subscript $S$ denotes the substrate. A similar equation holds for the interface between the film and the gate. From the continuity equation for the current follows $F^{e/h}(\epsilon)=x F_S^{e/h}(\epsilon)+(1-x)F^{e/h}_{\Gs}(\epsilon)$, where
$x={R_2}/({R_1+R_2})$.
Here $F_S^{e/h}(\epsilon)=\tanh\left({\epsilon}/2T_1\right)$ and $F_{\Gs}^{e/h}(\epsilon)=\tanh\left({(\epsilon\mp eV_{\Gs})}/2T_2\right)$ denote the distributions in the substrate and in the gate, respectively. The gauge invariant distribution in the film is defined as $\tilde{F}^{e/h}({\epsilon})=F^{e/h}({\epsilon\pm e\phi_{\cK}^{\mathrm{cl}}})$ and takes the form
\begin{align}\label{eq:distributioninthefilm}
\tilde{F}^{e/h}(\epsilon)=x \tanh{\left(\frac{\epsilon\pm (1-x)eV_{\Gs}}{2T_1}\right)} +(1-x) \tanh{\left(\frac{\epsilon\mp x eV_G}{2T_2}\right)}.
\end{align}
The upper (lower) signs correspond to electrons (holes).
We assumed very resistive interfaces, such that the resistance of the film can be neglected with respect to the resistance of the interfaces. Also, we assumed $V_G, T_1, T_2\ll \tau^{-1}$.

Now we can proceed with the evaluation of the expression (\ref{eq:noisenoninteracting}). In the case $T_1=T_2=T$ one can calculate it exactly and obtains
\begin{align}\label{eq:noiseEq}
S_0(\br-\br',\epsilon)=&S^{eq}_0(\br-\br',\epsilon)[x^2+(1-x)^2]
+S^{eq}_0(\br-\br',\epsilon+eV_G)x (1-x)
+S^{eq}_0(\br-\br',\epsilon-eV_G)x (1-x),
\end{align}
where $S^{eq}_0$ is the equilibrium noise, see Eq.~(\ref{eq:soeq}). When one of the tunneling resistances is infinitely large, then effectively the system is in contact only with one reservoir and therefore in an equilibrium. In that case $x=0$ or $x=1$, and as expected Eq.~(\ref{eq:noiseEq}) reduces to the equilibrium noise $S_0^{eq}$. Moreover, notice that by increasing the temperature of the system, thermal fluctuations increase and the current noise increases. Similarly, in Eq.~(\ref{eq:noiseEq}), the gate voltage also leads to an increase of the noise with respect to the equilibrium one ($V_G=0$). In the case of zero frequency, the noise $S_0(\br-\br',0)$ was found in Ref.~\onlinecite{PRL}. Eq.~(\ref{eq:noiseEq}) also describes a diffusive bridge placed between two reservoirs having different chemical potentials\cite{Altshuler}. There, $x$ plays the role of the coordinate along the bridge, and after performing the integration over $x$ in Eq.~(\ref{eq:noiseEq}) one finds the shot noise of that system.

Next we discuss the influence of the electron-electron interaction on the distribution function (\ref{eq:distributioninthefilm}). In general, the interaction leads to smearing of a noninteracting distribution function. However, when the inelastic length is large in comparison to the system dimensions, inelastic processes can be neglected. The inelastic length is expected to increase as the temperature and/or applied voltage decrease, allowing one to tune the ratio of the characteristic system length and the inelastic relaxation length. Coulomb-interaction induced corrections to the noise in the equilibrium have been studied by Altshuler and Aronov\cite{Aronov-Altshuler}. Out-of-equilibrium, the influence of the the Coulomb  interaction on the shot noise in diffusive contacts has been studied in the limit of large inelastic length in Refs.~\onlinecite{Beenakker,Nagaev,Gefen}, while contributions due to inelastic collisions have been studied in Ref.~\onlinecite{Kozub}. In the present paper, we assume that time-of-flight of the quasiparticles through the film is much smaller than the typical energy-relaxation time. We consider a thin film characterized by the large inelastic scattering length and focus on the influence of the BCS interaction on the longitudinal transport in the normal state.

\section{Superconducting fluctuations\label{sec:fluctuations1}}

Having discussed the noninteracting case in the previous section, we start the analysis of the influence of the BCS interaction on the current noise. The system is assumed to be in the normal state, but in the vicinity of the nonequilibrium superconducting transition. To be more precise, we do not discuss the Berezinskii-Kosterlitz-Thouless transition, but the crossover that in the equilibrium happens at the BCS critical temperature $T_c$. In the following we analyze this crossover in nonequilibrium conditions, studying influence of the gate voltage and temperatures of the reservoirs on the Cooper pair lifetime and the processes discussed in Sec.~\ref{sec:current-current}. In this and in the following sections we consider the case where the Ginzburg-Landau rate is much smaller than the effective temperature $T_e$, defined below by Eq.~(\ref{eq:Te}).

The saddle point equation of the action (\ref{eq:S}) for the superconducting order parameter has the trivial solution with the average value of the superconducting order parameter being zero. Then, one recovers the normal-metal saddle point (\ref{eq:saddle}) for the field $\check{Q}$. Therefore, now we include the massless fluctuations around it. We substitute Eq.~(\ref{parametrization}) into the action (\ref{eq:S}) and expand it to the second order in $\check{\mathcal{W}}$, Eq.~(\ref{eq:W}).  Since we are neither interested in the weak localization correction nor in the Altshuler-Aronov type corrections, but in fluctuations caused by the fluctuating superconducting order parameter $\Delta(\br,t)$, in the following we consider only the Cooperon degrees of freedom. After integrating out the Cooperon degrees of freedom, we obtain a Ginzburg-Landau-like action which depends only on the order parameter  $\Delta(\br,t)$. This calculation was performed in Ref.~\onlinecite{PRBmy} and the correlation functions of Cooperon fields have been obtained. They read as \footnote{Note that in Ref.~\onlinecite{PRBmy} it was used the notation $A(\bx)=\int\dif\bq A(\bq) \exp(i\bq\bx)/(2\pi)^d$ and $A^*(\bx)=\int\dif\bq A^*(\bq) \exp(i\bq\bx)/(2\pi)^d$ and similarly for the Fourier transform in the time-domain. This notation is confusing since then the convention is that complex conjugation includes the change of momentum and energy, $\epsilon\to-\epsilon$ and $\bq\to-\bq$. In the following, we will use $A(\bx)=\int\dif\bq A(\bq) \exp(i\bq\bx)/(2\pi)^d$ and $A^*(\bx)=\int\dif\bq A^*(-\bq)\exp(i\bq\bx)/(2\pi)^d$. },\cite{PRBmy}
\begin{widetext}
\begin{align}\label{corr1}
&\langle\langle
w_{\epsilon_1,\epsilon_2}({\bf{q}})w_{-\epsilon_3,-\epsilon_4}^*({\bf{q}})\rangle\rangle=\frac{2
i}{\nu}\delta_{\epsilon_1-\epsilon_2,\epsilon_4-\epsilon_3}\frac{-L^{-1}_K
L_{A,1-2}L_{R,1-2}+F^h(\epsilon_{3})L_{R,1-2}+F^e(\epsilon_1)L_{A,1-2}}{\left[D
\mathbf{q}^2-i(\epsilon_1+\epsilon_2)\right]\left[D \mathbf{q}^2-i (\epsilon_3+\epsilon_4)\right]},\\ \label{corr2}
&\langle\langle
\bar{w}_{\epsilon_1,\epsilon_2}({\bf{q}})\bar{w}_{-\epsilon_3,-\epsilon_4}^*({\bf{q}})\rangle \rangle=\frac{2
i}{\nu}\delta_{\epsilon_1-\epsilon_2,\epsilon_4-\epsilon_3}\frac{-L^{-1}_K
L_{A,1-2}L_{R,1-2}-F^h(\epsilon_{2})L_{A,1-2}-F^e(\epsilon_4)L_{R,1-2}}{\left[D
\mathbf{q}^2+i(\epsilon_1+\epsilon_2)\right]\left[D \mathbf{q}^2+i (\epsilon_3+\epsilon_4)\right]},
\\ \label{corr3}
&\langle\langle
\bar{w}_{\epsilon_1,\epsilon_2}({\bf{q}})w_{-\epsilon_3,-\epsilon_4}^*({\bf{q}})\rangle\rangle= \frac{2
i}{\nu}\delta_{\epsilon_1-\epsilon_2,\epsilon_4-\epsilon_3}\frac{L^{-1}_K
L_{A,1-2}L_{R,1-2}+F^h(\epsilon_2)L_{A,1-2}-F^h(\epsilon_3)L_{R,1-2}}{\left[D
\mathbf{q}^2+i(\epsilon_1+\epsilon_2)\right]\left[D \mathbf{q}^2-i (\epsilon_3+\epsilon_4)\right]}.
\end{align}
\end{widetext}
The average $\langle\langle\ldots\rangle\rangle$ is with respect to the action $S$ given by Eq.~(\ref{eq:S}) and it includes averaging over the fluctuations of $\check{Q}$, $\Delta^{cl}$ and $\Delta^{q}$. Here  $L_{R/A,i-j}\equiv\left(L_{R/A}^{-1}(\bq,\epsilon_i-\epsilon_j) \right)^{-1}$ denotes retarded/advanced fluctuation propagators and $L_K^{-1}$ is the Keldysh propagator. Their low frequency $\omega$ and low momentum $q$ behavior is given by the following expressions\cite{PRBmy}:
\begin{align}\label{eq:Lk}
L^{-1}_{K}&=i\frac{\pi}{2}\left[1-\tilde{F}^h(0)\tilde{F}^e(0)\right] ,\\\label{eq:L_R}
L^{-1}_{R/A}(\bq,\epsilon)&=\frac{\pi}{8 T_e}\Big\{-(\tau_{\mathrm{\GL}}z_{\cp})^{-1}+
\Big[\mp 4iT_e \tilde{F}^R(0)-D\bq^2\pm i\epsilon\mp 2ieV_G(1-x)
\Big] \Big(1\pm i\frac{T_e}{\Omega}\Big)\Big\}.
\end{align}
Here and in Eqs.~(\ref{corr1}-\ref{corr3}) we set $A_{\cK}=0$.
The parameters appearing in the retarded and advanced propagators are functionals of
$ F^R({\epsilon})=\left[F^h({\epsilon})-F^e({-\epsilon})\right]/2$.
They are given by the following expressions $T_e^{-1}=2\partial_{\epsilon}\tilde{F}^R({\epsilon})\Big|_{\epsilon=0}$, $\Omega^{-1}={2}\dashint \dif \epsilon \left[{\tilde{F}^R({\epsilon})-\tilde{F}^R({0})}\right]/({\epsilon^2{\pi}})$,
and $z_{\cp}^{-1}=1+\left({T_e}/{\Omega}\right)^2$.
The symbol $\dashint$ denotes the principal value of the integral. The nonequilibrium Ginzburg-Landau (GL) rate is defined as
$
\tau_{\mathrm{\GL}}^{-1}=-{4}z_{\cp}T_e\int_{-\omega_D}^{+\omega_D} \dif\epsilon \left[\tilde{F}^R({\epsilon})- \tanh{\left(\epsilon/2T_c\right)}\right]/\pi\epsilon+4z_{\cp}{T_e^2}{\Omega^{-1}}\tilde{F}^R(0)
$,
where $\omega_D$ is the Debye energy. The GL time denotes the lifetime of the fluctuation induced Cooper pairs. It carries the information how far the system is from the transition and at the transition \footnote{More precisely, in two dimensions this condition does not correspond to a real superconducting transition, but to a crossover.} it becomes infinitely large, signaling that the Cooper pairs become long living. We point out that the fluctuation propagators given above are valid when the system is in normal state but in the vicinity of the transition such that $(\tau_{\mathrm{\GL}}T_e)^{-1}\ll 1$. Substituting the distribution function (\ref{eq:distributioninthefilm}) in the previous equations, one evaluates all these parameters and finds that they are functions of temperatures of the reservoirs $T_1$ and $T_2$, the ratio of tunneling resistances $x=R_2/(R_1+R_2)$ and the gate voltage $V_G$\cite{PRBmy}:
\begin{align}
\label{eq:Te}
T_\mathrm{e}=&\left[\frac {x}{T_1\cosh^2\frac {(1-x)e V_{\rm\scriptscriptstyle G}}{2T_1}}
+\frac {(1-x)}{T_2\cosh^2\frac {x e V_{\rm\scriptscriptstyle G}}{2T_2}}\right]^{-1},
\\
\label{eq:invOmega}
\Omega^{-1}=&\frac{2x}{T_1 \pi^2}\mathrm{Im}\left[ \Psi '\left(\frac{1}{2}-i\frac{eV_{\Gs} (1-x)}{2\pi T_1}\right)\right]+\frac{2(1-x)}{T_2 \pi^2}  \mathrm{Im}\left[\Psi'\left(\frac{1}{2}+i\frac{eV_{\Gs} x}{2\pi T_2}\right)\right],
\\ \label{eq:tGL}\tau_{\mathrm{\GL}}^{-1}=&\frac{8}{\pi}z_{\cp}T_e\Bigg\{ x\mathrm{Re}\left[\Psi\left(\frac{1}{2}+i\frac{(1-x)eV_{\Gs}}{2\pi T_1}\right)\right] +(1-x)\mathrm{Re}\left[\Psi\left(\frac{1}{2}+i\frac{xeV_{\Gs}}{2\pi T_2}\right)\right]+2\ln2\notag\\&+x \ln\frac{T_1}{T_2}-\ln\frac{T_c}{T_2}+\gamma \Bigg\}+4z_{\cp}\frac{T_e^2}{\Omega}\tilde{F}^R(0).
\end{align}
Here $\Psi(z)$ is the digamma function, defined as  $\Psi(x)=\Gamma'(x)/\Gamma(x)$, where $\Gamma(x)$ is the gamma function. We expressed the critical temperature as $T_c=2\omega_D \exp{(\gamma-\lambda^{-1})}/\pi$, where $\gamma$ is the Euler constant. Now one easily finds the parameter $z_{\cp}^{-1}=1+\left({T_e}/{\Omega}\right)^2$.

We are now equipped to start the calculation of different fluctuation contributions to the current-current correlation function. However, before doing it, we make a brief digression in order to stress the importance of the above given fluctuation propagators. In close-to-equilibrium conditions, the Aslamazov-Larkin contribution to the conductivity can be obtained by employing together the linear response theory and the phenomenological time-dependent GL (TDGL) equation, while the Maki-Thompson and the DOS, should be derived starting from the microscopic theory. The phenomenological TDGL equation reads as
\begin{align}\label{eq:tdglphenom}
\left[-\frac{8}{\pi}(T-T_c)-D(\bq-2e\mathbf{\Ak})^2+i(\omega-2e\phi_{\mathcal{K}})\right]
\Delta^{cl}_{\mathcal{K}}(\bq,\omega)+\zeta(\bq,\omega)=0,
\end{align}
where the thermal noise satisfies
\begin{align} \label{eq:noise}
\langle \zeta(\br,t)\zeta^*(\br',t')\rangle=\frac{16}{\pi\nu} T^2\delta(\br-\br')\delta(t-t').
\end{align}
However, if one is interested in a nonlinear dependence on the drive, then the previous equation is not a good starting point. Notice that even for the simple setup here considered (see Fig.~\ref{fig:film}), the TDGL equation takes different form\cite{PRBmy} than the usual phenomenological TDGL equation (\ref{eq:tdglphenom}). It is given by
 \begin{align}\label{eq:TDGL}
\frac{8T_e}{\pi}L_R^{-1}(\bq,\omega)\Delta^{cl}_{\mathcal{K}}(\bq,\omega)
+\zeta(\bq,\omega)=0,
\end{align}
 where the nonequilibrium noise satisfies the following condition
\begin{align}\label{eq:noiseNEQ}
 \langle \zeta(\br,t)\zeta^*(\br',t')\rangle=&\frac{16}{\pi\nu} T_e^2\left[1-\tilde{F}^h(0)\tilde{F}^e(0)\right]\delta(\br-\br')\delta(t-t').
\end{align}
In Eq.~(\ref{eq:TDGL}), the retarded fluctuation propagator is given by Eq.~(\ref{eq:L_R}) with many drive-dependent parameters that we discussed above. Only in the equilibrium, $V_G=0$ and $T_1=T_2$, Eqs.~(\ref{eq:TDGL}) and (\ref{eq:noiseNEQ}) coincide with Eqs.~(\ref{eq:tdglphenom}) and (\ref{eq:noise}), while otherwise they are different. Therefore, we point out that in order to study a nonlinear dependence of some observable on the drive, one should derive a corresponding TDGL equation starting from the microscopic theory and not to use the phenomenological one (\ref{eq:tdglphenom}), as is usually the case in literature.

\section{Density of states contribution\label{sec:DOS}}

In this section we start the calculation of the current correlation function (\ref{eq:noisedef}) in the presence of the BCS interaction. We consider linear response to the in-plane electric field, while the system is driven out-of-equilibrium by thermal baths having different temperatures, between which it is sandwiched, or by an electric field perpendicular to the film, see Fig.~\ref{fig:film}.

First we consider the density of state contribution to the noise.  Taking into account the massless fluctuations around the normal-metal saddle point Eq.~(\ref{eq:saddle}) to the second order in the Cooperon fields, we collect all the terms of the type  $\langle\langle\bar{w}\bar{w}^*\rangle\rangle$ and $\langle\langle{w}{w}^*\rangle\rangle$ in Eq.~(\ref{eq:noisedef}). They constitute the DOS contribution and together give
\begin{widetext}
\begin{align}\label{eq:DOS}
S_{\DOS}(\br-\br',\epsilon)=&-\frac{\nu D e^2}{16\pi}\delta(\br-\br')\int\dif\epsilon_1\dif\epsilon_2\Bigg\{
\langle\langle \bar{w}_{\epsilon_1,\epsilon_2}(\br)\bar{w}^*_{-\epsilon_2,-\epsilon_1}(\br)+
{w}_{\epsilon_1,\epsilon_2}(\br){w}^*_{-\epsilon_2,-\epsilon_1}(\br)\rangle\rangle
\notag\\&\times\left[ 1-F^h(\epsilon_2-\epsilon)F^h(\epsilon_2)
-F^e(\epsilon_1-\epsilon)F^e(\epsilon_1)\right]
-\langle\langle w_{\epsilon_1,\epsilon_2}(\br)w^{*}_{-\epsilon_2-\epsilon,-\epsilon_1-\epsilon}(\br)
+\bar{w}^{*}_{-\epsilon_2,-\epsilon_1}(\br)
\bar{w}_{\epsilon_1+\epsilon,\epsilon_2+\epsilon}(\br)\rangle\rangle\notag\\&\times F^h(\epsilon_2)F^e(\epsilon_1+\epsilon)\Bigg\}+\epsilon\to-\epsilon.
\end{align}
The leading contribution in Eq.~(\ref{eq:DOS}) close to the transition into superconducting state is given by
\begin{align}\label{eq:mainDOS}
S_{\DOS}(\br-\br',\epsilon)=&+\delta(\br-\br')\frac{D e^2}{16\pi^3d_f}\int\dif^2\bq \dif E \dif \omega  \mathrm{Im}\Bigg(L_K^{-1}L_{A,\omega}L_{R,\omega}\bigg\{
2\frac{F^h(E-\epsilon)F^h(E)}{[Dq^2-i(2E+\omega)]^2}\notag\\&+
\frac{F^h(E-\omega)F^e(E+\epsilon)}{[Dq^2-i(2E+2\epsilon-\omega)][Dq^2-i(2E-\omega)]}
\bigg\}\Bigg)
+\epsilon\to-\epsilon.
\end{align}
We use the notation  $L_{R/A,\omega}\equiv\left(L_{R/A}^{-1}(\bq,\omega) \right)^{-1}$. Notice that Eq.~(\ref{eq:mainDOS}) is the only singular part of Eq.~(\ref{eq:DOS}) for $\tau_{\GL}^{-1}\to 0$ and therefore the most dominant close to the transition. The holes and electrons give the same contribution. This is expected, since the system is in the normal state and there is no long-living Cooper-pair condensate that could allow for a charge imbalance between the holes and electrons. We further evaluate Eq.~(\ref{eq:mainDOS}) and find up to logarithmic accuracy:
\begin{align}\label{eq:dosfinal}
S_{\DOS}(\br-\br',\epsilon)\approx&-\frac{e^2}{2\pi^2d_f}\delta(\br-\br')
\left[1-\tilde{F}^h(0)\tilde{F}^e(0)\right] T_e^2 z_{cp} \ln{\left(\frac{T_e}{\tau_{\GL}^{-1}}\right)}
\mathrm{Re}\Bigg\{2\int\dif E\frac{\tilde{F}^h(E-\epsilon)\tilde{F}^h(E)}
{(E+i0)^2}\notag\\&+
\int\dif E\frac{\tilde{F}^h(E)\tilde{F}^e(E+\epsilon)}{(E+\epsilon+i0)
(E+i0)}\Bigg\}+\epsilon\to-\epsilon.
\end{align}
\end{widetext}
Here and in the following $i0$ denotes $i0^+$. In order to discuss the assumptions made in Eq.~(\ref{eq:dosfinal}), we introduce
\begin{align}\label{eq:epsilono}
\epsilon_0=-4T_e\tilde{F}^R(0)+T_e\tau_{\GL}^{-1}/{\Omega}.
\end{align}
We assume that the system is close to the transition such that $|\epsilon_0|$ and $\tau_{\GL}^{-1}$ are much smaller than relevant energy scales of the distribution function, i.e., $\mathrm{min}(V_G,T_1,T_2)$. For the special case $x=1/2$ and $T_1=T_2$, $\epsilon_0$ is exactly zero. Eq.~(\ref{eq:dosfinal}) is also valid for $V_G=0$ and then the condition becomes $|\epsilon_0|,\tau_{\GL}^{-1}\ll \mathrm{min}(T_1,T_2)$. When obtaining Eq.~(\ref{eq:dosfinal}), we used the fact that the most important contribution in Eq.~(\ref{eq:mainDOS}) originates from small momenta and that therefore we can safely cut the momentum integration at the upper limit $Dq^2_{max}\sim T_e$.

In the case of equal temperatures of the reservoirs $T_1=T_2=T$, substituting the distribution function (\ref{eq:distributioninthefilm}) in Eq.~(\ref{eq:dosfinal}), we obtain the nonlinear dependence of the noise on the gate voltage $V_G$, temperature $T$ and frequency $\epsilon$. Integrating out spatial coordinates, one obtains the final result
\begin{align}\label{eq:dosresult}
S_{\DOS}(\epsilon)\approx&-\frac{e^2}{\pi^3d_f}
\left[1-\tilde{F}^h(0)\tilde{F}^e(0)\right] T_e^2 z_{cp} \ln{\left(\frac{T_e}{\tau_{\GL}^{-1}}\right)} \Big[
x^2D\left(eV_G(1-x),eV_G(1-x)\right)\notag\\&
+x(1-x)D\left(eV_G(1-x),-eV_Gx\right)
+x(1-x)D\left(-eV_Gx,eV_G(1-x)\right)\notag\\&
+(1-x)^2D\left(-eV_Gx,-eV_Gx\right)+\epsilon\to-\epsilon\Big]
\end{align}
The function $D(x,y)$ is defined as
\begin{widetext}
\begin{align}\label{eq:eqD}
D(x,y)=&\frac{1}{T}\coth{\left(\frac{y-x+\epsilon}{2T}\right)}\mathrm{Im}\left[
\Psi'\left(\frac{1}{2}-i\frac{\epsilon-x}{2\pi T}\right)-\Psi'\left(\frac{1}{2}+i\frac{y}{2\pi T}\right)\right]
-\frac{\pi}{\epsilon}\coth{\left(\frac{x+y+\epsilon}{2T}\right)}
\notag\\&\times\mathrm{Re}\left[
\Psi\left(\frac{1}{2}-i\frac{x}{2\pi T}\right)-\Psi\left(\frac{1}{2}-i\frac{x+\epsilon}{2\pi T}\right)
+\Psi\left(\frac{1}{2}+i\frac{y}{2\pi T}\right)
-\Psi\left(\frac{1}{2}+i\frac{y+\epsilon}{2\pi T}\right)
\right],
\end{align}
\end{widetext}
where the frequency satisfies $\epsilon \gg |\epsilon_0|, \tau_{\GL}^{-1}$. However, for $\epsilon\ll\mathrm{min}(V_G,T_1,T_2)$ but arbitrary with respect to $|\epsilon_0|, \tau_{\GL}^{-1}$, one obtains the result by setting $\epsilon$ to zero in Eq.~(\ref{eq:dosresult}). The expressions for $\tau_{\GL}$, $T_e$, $z_{cp}$ and $\Omega$ are provided in Sec.~\ref{sec:fluctuations1}.  In Appendix~\ref{app:B}
we discuss the case of different temperatures of the reservoirs,  $T_1\neq T_2$.

Next we analyze the expression (\ref{eq:dosresult}). For $x=0$ (or $x=1$) the system is effectively in an equilibrium and Eq.~(\ref{eq:dosresult}) simplifies to
\begin{align}\label{eq:noiseDoseq}
S_{\DOS}^{eq}(\epsilon)=&-\frac{2e^2 T}{\pi^3d_f}\ln{\left(\frac{T}{\tau_{\GL}^{-1}}\right)}
\coth{\left(\frac{\epsilon}{2T}\right)}\Bigg\{\mathrm{Im}\left[\Psi'\left( \frac{1}{2}-i\frac{\epsilon}{2\pi T}\right)\right]
-2\pi\frac{T}{\epsilon}\mathrm{Re}\left[\Psi\left(\frac{1}{2}\right)-
\Psi\left(\frac{1}{2}-i\frac{\epsilon}{2\pi T}\right)\right]\Bigg\}.
\end{align}
For $V_G=0$ we also get the same expression from Eq.~(\ref{eq:dosresult}), since it also corresponds to the equilibrium situation. In the case of noninteracting equilibrium electrons, studied in Sec.~\ref{sec:noninteracting}, the source of the current fluctuations is thermal fluctuations. Here we obtained the additional contribution (\ref{eq:dosresult}) to the current noise caused by superconducting fluctuations. The Cooper pair density fluctuates due to randomness in the formation and dissociation of Cooper pairs, and these fluctuations affect the single-particle density of states leading to the noise $S_{\DOS}$.

\begin{figure}
\includegraphics[width=0.55\columnwidth]{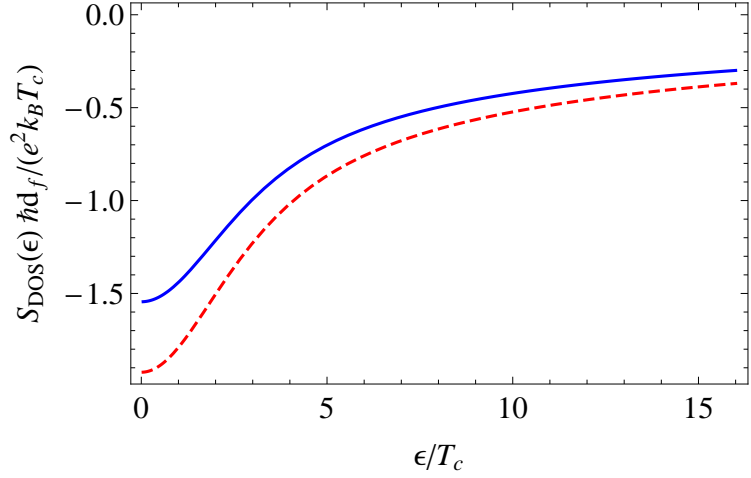}\\
\caption{Current noise $S_{\DOS}(\epsilon)$ is shown as a function of frequency $\epsilon$ for the case of equal temperatures of the reservoirs $T_1=T_2=1.01 T_c$, the ratio of tunneling resistances $x=1/3$. The red dashed line presents zero gate voltage and the blue solid line $V_G=T_c/2$. \label{fig:DOS}}
\end{figure}

In equilibrium the FDT holds and relates the current fluctuations (noise) to the real part of the ac conductivity that provides information about absorbed energy in the sample:
\begin{align}
S^{eq}(\epsilon)=\mathrm{Re}[\sigma(\epsilon)] \epsilon \coth{\left(\frac{\epsilon}{2T}\right)},
\end{align}
where $S(\epsilon)=\int\dif \bx S(\bx,\epsilon)$. Using it we obtain the density of states contribution to the in-plane ac conductivity in equilibrium:
\begin{align}\label{eq:ac}
\mathrm{Re}[\sigma_{\DOS}^{eq}(\epsilon)]=&-\frac{2e^2 }{\pi^3d_f}\frac{T}{\epsilon}\ln{\left(\frac{T}{\tau_{\GL}^{-1}}\right)}
\Bigg\{\mathrm{Im}\left[\Psi'\left( \frac{1}{2}-i\frac{\epsilon}{2\pi T}\right)\right]
-2\pi\frac{T}{\epsilon}\mathrm{Re}\left[\Psi\left(\frac{1}{2}\right)-
\Psi\left(\frac{1}{2}-i\frac{\epsilon}{2\pi T}\right)\right]\Bigg\}\\
 \label{eq:limitDOS}\approx&
\begin{cases} -\frac{21 e^2 \zeta(3)}{\pi^4d_f}\ln{\left(\frac{T}{\tau_{\GL}^{-1}}\right)}, & \epsilon\ll T\\
-\frac{4e^2}{\pi^2d_f} \ln{\left(\frac{T}{\tau_{\GL}^{-1}}\right)}\left(\frac{T}{\epsilon}\right)^2 \ln{\left(\frac{\epsilon}{T}\right)},& \epsilon\gg T. \end{cases}
 \end{align}
This result is in the agreement with findings of Ref.~\onlinecite{ACmy}, where it is obtained in a different manner, i.e.~directly considering the ac conductivity in equilibrium within the nonlinear sigma model approach. In Eqs.~(\ref{eq:noiseDoseq},\ref{eq:ac},\ref{eq:limitDOS}) the GL rate takes its equilibrium value $\tau_{\GL}^{-1}=8 T\ln{(T/T_c)}/\pi$. One obtains that suppression of the single-particle density of states due to superconducting fluctuations gives the negative contribution to the Drude conductivity, see Fig.~\ref{fig:DOS} where  $S^{eq}_{\DOS}$ as the function of frequency $\epsilon$ is shown by the red dashed line. We remind the reader that our calculations apply when the frequency, temperatures and the gate voltage are much smaller than the elastic scattering rate, since we use the nonlinear sigma model that captures low-energy physics.

Now we consider the influence of the gate voltage on the noise $S_{\DOS}$. From Eq.~(\ref{eq:dosresult}) one easily finds that the first correction to the equilibrium noise (\ref{eq:noiseDoseq}) in the limit $V_G\ll T$ depends quadratically on the gate voltage. This is expected since in the case of equal temperatures of the reservoirs, $T_1=T_2=T$, in-plane current is invariant under the transformation $V_G\to -V_G$ and therefore $S_{\DOS}$,  Eq.~(\ref{eq:dosresult}), is the even function of the gate voltage.  Fig.~\ref{fig:DOS} shows $S_{\DOS}$, given by Eq.~(\ref{eq:dosresult}), as a function of $\epsilon$ for the case $T_1=T_2=1.01 T_c$ and the ratio of tunneling resistances $x=1/3$. The red dashed line represents zero gate voltage, i.e.~the equilibrium case, while the solid blue line is for $V_G=T_c/2$.  One observes that the gate voltage tends to suppress (the absolute value of) the equilibrium noise, $S_{\DOS}^{eq}$, contrary to the noninteracting electron case where it increases the noise, see Sec.~\ref{sec:noninteracting}. The reason is that the gate voltage decreases the Cooper pair lifetime and their density, and therefore the influence of superconducting fluctuations on the current correlation function decreases. The lifetime is given by the GL time $\tau_{\GL}$, see Eq.~(\ref{eq:tGL}).

\section{Maki-Thompson contribution\label{sec:MT}}

In this section we consider the Maki-Thompson contribution to the current-current correlation function. Its physical origin is already discussed in Sec.~\ref{sec:NLSM} and here we calculate it. We start from Eq.~(\ref{eq:noisedef}) and collecting all the terms that contain the average of the convolution of  $w$ and $\bar{w}^*$ fields, we find the the Maki-Thompson contribution
\begin{widetext}
\begin{align}\label{eq:MT}
  S_{\MT}(\br-\br',\epsilon)=&\frac{\nu D e^2}{16\pi}\delta(\br-\br')\int \dif E \dif\omega
   \Big\{ \langle \langle w_{E+\omega,E}(\br)\bar{w}^*_{-E-\epsilon,-E-\epsilon-\omega}(\br)\rangle\rangle
   \left[-1+F^h(E)F^h(E+\epsilon)\right]
   \notag\\&
  +\langle \langle \bar{w}_{E+\omega,E}(\br){w}^*_{-E-\epsilon,-E-\epsilon-\omega}(\br)\rangle\rangle
   \left[-1+F^e(E+\omega)
  F^e(E+\epsilon+\omega)\right]
  + \epsilon\to -\epsilon
 \Big\}.
\end{align}
The leading contribution of the previous expression close to the superconducting transition reads as
\begin{align}\label{eq:MTan}
S_{\MT}(\epsilon)=&\frac{e^2}{2\pi d_f}z_{cp}\left[1-\tilde{F}^h(0)\tilde{F}^e(0)\right] T_e^2\frac{\pi|\epsilon|+2\tau_{\GL}^{-1}\ln{\left( \frac{\tau_{\GL}^{-1}}{|\epsilon|}\right)}}{\tau_{\GL}^{-2}+\epsilon^2}\Big\{2-
\tilde{F}^{h}(0)
[\tilde{F}^{h}(\epsilon)
+\tilde{F}^{h}(-\epsilon)]\Big\},
\end{align}
\end{widetext}
and is valid for i) $\epsilon\ll \mathrm{min}\{T_1,T_2,V_G\}$ or ii) $V_G=0$ and $\epsilon\ll \mathrm{min}\{T_1,T_2\}$, but arbitrary with respect to $\tau_{\GL}^{-1}$. The condition for $\tau_{\GL}$ and $\epsilon_0$, Eq.~(\ref{eq:epsilono}), in comparison with the temperatures and the gate voltage is the same as for the density of states correction, see the discussion below Eq.~(\ref{eq:epsilono}). For the definitions of the parameters appearing in Eq.~(\ref{eq:MTan}), see Sec.~\ref{sec:fluctuations1}, while the distribution function is given by Eq.~(\ref{eq:distributioninthefilm}). One sees that $S_{\MT}$ is not singular function of the GL rate at finite frequency $\epsilon$. In the case $\epsilon\ll\tau_{\phi}^{-1}\ll\tau_{\GL}^{-1}$, the phase breaking rate $\tau_{\phi}^{-1}$ appears in Eq.~(\ref{eq:MTan}) inside the logarithm instead of $\epsilon$, because it serves as a cutoff scale at small momentum: $Dq^2_{\mathrm{\min}}\sim\tau_{\phi}^{-1}$.

\begin{figure}
\includegraphics[width=0.55\columnwidth]{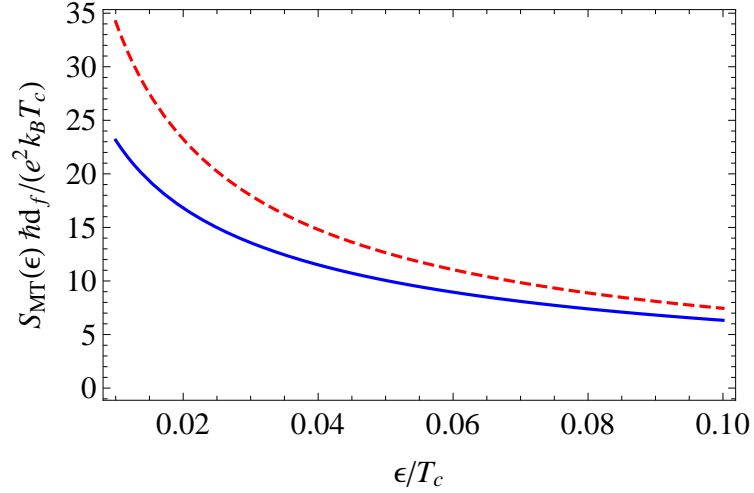}\\
\caption{Current noise $S_{\MT}(\epsilon)$ as a function of the frequency $\epsilon\gg\tau_{\phi}^{-1}$ for the case of equal temperatures of the reservoirs $T_1=T_2=1.01 T_c$ and the ratio of tunneling resistances $x=1/3$. The red dashed line presents zero gate voltage and the blue solid line $V_G=T_c/2$. \label{fig:MTan}}
\end{figure}

Note that contrary to the density of states contribution where the characteristic frequencies are determined by the temperature and the gate voltage, here two additional energy scales appear: the GL rate and the phase-breaking rate. Fig.~\ref{fig:MTan} shows $S_{\MT}$ as a function of frequency $\epsilon$ for the case $T_1=T_2=1.01 T_c$ and the ratio of tunneling resistances $x=1/3$. It is assumed that $\epsilon$ is greater than the phase-breaking rate. The red dashed line presents equilibrium case (zero gate voltage) and the blue solid line corresponds to $V_G=T_c/2$. We see that $S_{\MT}$  monotonically decreases with frequency, and is positive contrary to the negative  $S_{\DOS}$ contribution. On the other hand, the role of the gate voltage is similar in both $S_{\DOS}$ and $S_{\MT}$, it suppresses superconducting
fluctuations and therefore their influence on the current noise.

\begin{figure}\includegraphics[width=0.58\columnwidth]{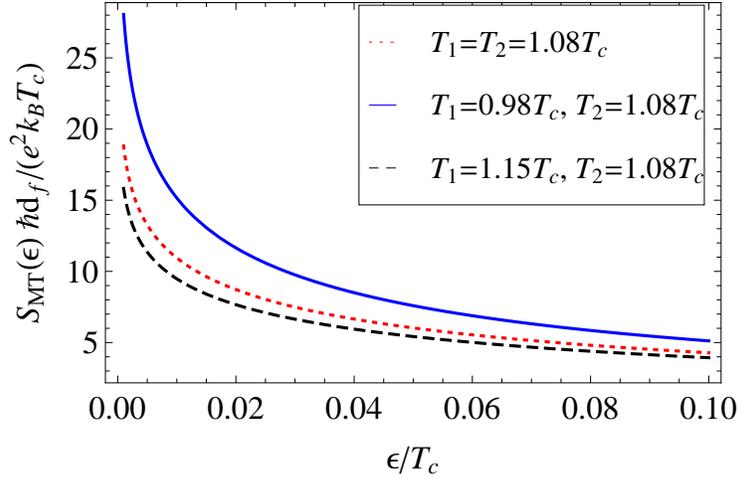}\\
\caption{Current-current correlation function $S_{\MT}$ as a function of frequency $\epsilon$ for different temperature of the baths. The ratio of tunneling resistances is fixed to be $x=1/3$ and the gate voltage is $V_G=0$.
\label{fig:MTdifT}}
\end{figure}

Further, we discuss the influence of temperatures of the thermal baths on the noise. For simplicity we consider the zero gate voltage. The system is driven out-of-equilibrium due to different temperatures of the reservoirs. Fig.~\ref{fig:MTdifT} shows frequency dependence of  $S_{\MT}$ for different realizations. The ratio of tunneling resistances is fixed to be $x=1/3$. The red dotted curve corresponds to the equilibrium situation $T_1=T_2=1.08 T_c$.  Lowering one of the temperatures,  the effective temperature of the system decreases, the lifetime of the Cooper pairs increases and therefore the noise also increases, as can be seen by comparing the blue solid curve in Fig.~\ref{fig:MTdifT}, that corresponds to the experimental realization $T_1=0.98T_c$ and $T_2=1.08 T_c$, with the red dotted curve. On the other hand, by increasing one of the temperatures, the effective temperature of the system increases, the lifetime of Cooper pairs decreases and therefore the influence of the superconducting fluctuations decreases, leading to the reduction of the noise, as shown by the black dashed curve for the case $T_1=1.15T_c$ and $T_2=1.08 T_c$.

In equilibrium, Eq.~(\ref{eq:MTan}) further simplifies. Using the FDT we find the leading MT ac conductivity\cite{acMT}
\begin{align}\label{eq:EqMTan}
\mathrm{Re}[\sigma_{\MT}(\epsilon)]=& \frac{e^2}{2\pi d_f} T\frac{\pi|\epsilon|+2\tau_{\GL}^{-1}\ln{\left( \frac{\tau_{\GL}^{-1}}{|\epsilon|}\right)}}{\tau_{\GL}^{-2}+\epsilon^2},\\
\approx &
\begin{cases}
\frac{e^2}{\pi d_f}\frac{T}{\tau_{\GL}^{-1}}\ln{\left(\tau_{\GL}^{-1}/|\epsilon|\right)}, & T\gg\tau_{\GL}^{-1}\gg \epsilon\\
\frac{e^2}{2 d_f}\frac{T}{|\epsilon|},& T\gg\epsilon\gg  \tau_{\GL}^{-1}.
\end{cases}
\end{align}
Here the GL rate takes its equilibrium value $\tau_{\GL}^{-1}=8 T\ln{(T/T_c)}/\pi$ and $\tau_{\phi}^{-1}\ll\epsilon\ll T$. In the case $\epsilon\ll\tau_{\phi}^{-1}\ll\tau_{\GL}^{-1}$, the phase breaking rate $\tau_{\phi}^{-1}$ appears inside the logarithm instead of $\epsilon$, because it serves as a cutoff scale at small momentum: $Dq^2_{\mathrm{\min}}\sim\tau_{\phi}^{-1}$.

\section{Aslamazov-Larkin contribution\label{sec:AL}}

In this section we obtain and analyze the Aslamazov-Larkin contribution to the current-current correlation function. Having discussed the density of states and the Maki-Thompson terms in Eq.~(\ref{eq:noisedef}), all the remaining terms are the fourth order in the Cooperon degrees of freedom. They constitute the Aslamazov-Larkin contribution.  We use the notation $S_{\AL}(\epsilon)=\int \dif \br S_{\AL}(\br,\epsilon)$ and find
\begin{widetext}
\begin{align}\label{eq:AL}
S_{\AL}(\epsilon)=&-\frac{1}{d_f}\left(\frac{\pi \nu e D}{2}\right)^2\int _{\epsilon_1,\epsilon_2,\epsilon_4,\epsilon_5,\bq_1,\bq_3} q_{1,x}q_{3,x}\Big\langle\Big\langle \Big[ -F^h(\epsilon_1) \bar{w}_{\epsilon_2,\epsilon_1-\epsilon}(\bq_1)
\bar{w}^*_{-\epsilon_1,-\epsilon_2}(\bq_1)\notag\\&+
F^e(\epsilon_1)\bar{w}^*_{-\epsilon_2,-\epsilon_1+\epsilon}(-\bq_1)
\bar{w}_{\epsilon_1,\epsilon_2}(-\bq_1)
+F^h(\epsilon_1-\epsilon) {w}_{\epsilon_2,\epsilon_1-\epsilon}(\bq_1)
{w}^*_{-\epsilon_1,-\epsilon_2}(\bq_1)\notag\\&-
F^e(\epsilon_1-\epsilon){w}^*_{-\epsilon_2,-\epsilon_1+\epsilon}(-\bq_1)
{w}_{\epsilon_1,\epsilon_2}(-\bq_1)
\Big]
\Big[ -F^h(\epsilon_4) \bar{w}_{\epsilon_5,\epsilon_4+\epsilon}(\bq_3)
\bar{w}^*_{-\epsilon_4,-\epsilon_5}(\bq_3)\notag\\&+
F^e(\epsilon_4)\bar{w}^*_{-\epsilon_5,-\epsilon_4-\epsilon}(-\bq_3)
\bar{w}_{\epsilon_4,\epsilon_5}(-\bq_3) +F^h(\epsilon_4+\epsilon) {w}_{\epsilon_5,\epsilon_4+\epsilon}(\bq_3)
{w}^*_{-\epsilon_4,-\epsilon_5}(\bq_3)\notag\\
&-
F^e(\epsilon_4+\epsilon){w}^*_{-\epsilon_5,-\epsilon_4-\epsilon}(-\bq_3)
{w}_{\epsilon_4,\epsilon_5}(-\bq_3)\Big]\Big\rangle\Big\rangle.
\end{align}
where $\int_{\epsilon_i}\equiv\int\dif\epsilon_i/(2\pi)$ and $\int_{\bq_i}\equiv\int\dif \bq_i/(2\pi)^2$.
The leading contribution in Eq.~(\ref{eq:AL}) close to the transition into the superconducting state is given by
\begin{align}\label{eq:56}
S_{\AL}(\epsilon)=&\frac{1}{d_f}\left\{\frac{16eDT_e^2z_{cp}}{\pi^2}
\left[1-\tilde{F}^h(0)\tilde{F}^e(0)\right]\right\}^2
\int_{-\infty}^{+\infty}\dif\omega\int_{0}^{+\infty}\dif q q^3\frac{1}{(\tau^{-1}_{\GL}+Dq^2)^2+(\epsilon-\omega)^2}
\frac{1}{(\tau^{-1}_{\GL}+Dq^2)^2+\omega^2}\notag\\
&\times\left(\left\{\mathrm{Im}\left[\int \dif E\frac{\tilde{F}^h(E)}{(2E-i0)^2}\right]\right\}^2+V_{G}\to -V_{G}\right).
\end{align}
The condition of applicability of the last formula is the same as for the density of states correction, see the discussion below Eq.~(\ref{eq:epsilono}).
Since we used the fluctuational propagators from Sec.~\ref{sec:fluctuations1} that are applicable for low frequencies, Eq.~(\ref{eq:56}) is valid for $\epsilon$ much smaller than the reservoir temperatures and the gate voltage. However, Eq.~(\ref{eq:56}) is valid also for $V_G=0$ and then the condition is $\epsilon\ll\mathrm{min}(T_1,T_2)$.
Evaluating Eq.~(\ref{eq:56}), we find
\begin{align}\label{eq:finalAL}
S_{\AL}(\epsilon)=&\frac{8}{\pi^5d_f}\left\{e z_{cp} T_e^2\left[1-\tilde{F}^h(0)\tilde{F}^e(0)\right]\right\}^2\left[
\frac{1}{\epsilon}\arctan{\left(\frac{\epsilon}{2\tau_{\GL}^{-1}}\right)}-
\frac{\tau_{\GL}^{-1}}{\epsilon^2}\ln{\left(1+\frac{\epsilon^2}{4\tau_{\GL}^{-2}}\right)}
\right]\notag\\ &\times \left\{\left\{\frac{x}{T_1}\mathrm{Re}\left[\Psi'\left(\frac{1}{2}+\frac{i}{2\pi T_1}(1-x)eV_G\right)\right] +\frac{1-x}{T_2}\mathrm{Re}\left[\Psi'\left(\frac{1}{2}-\frac{i}{2\pi T_2}xeV_G\right)\right]\right\}^2+V_G\to-V_G\right\}.
\end{align}
\end{widetext}
For the definitions of $z_{cp}$, $\tau_{\GL}^{-1}$ and $T_e$ see Sec.\ref{sec:fluctuations1}. The distribution of electrons and holes in the film is given by Eq.~(\ref{eq:distributioninthefilm}), where $x=R_2/(R_1+R_2)$ is ratio of tunneling resistances, and $d_f$ is the film thickness.
\begin{figure}
\includegraphics[width=0.55\columnwidth]{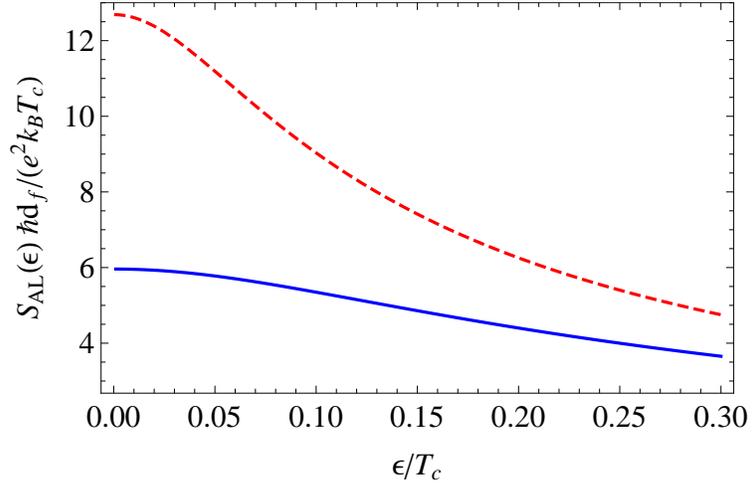}
\caption{Current noise $S_{\AL}(\epsilon)$ as a function of frequency $\epsilon$ for the case of equal temperatures of the reservoirs $T_1=T_2=1.01 T_c$ and the ratio of tunneling resistances $x=1/3$. The red dashed line presents zero gate voltage and the blue solid line $V_G=T_c/2$. \label{fig:AL}}
\end{figure}

\begin{figure}\includegraphics[width=0.65\columnwidth]{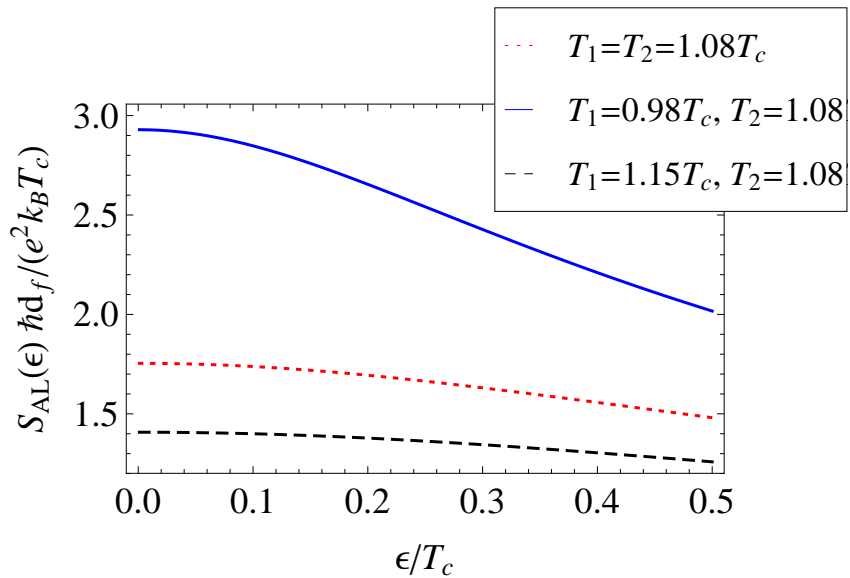}\\
\caption{Current-current correlation function $S_{\AL}$ as a function of frequency $\epsilon$ for different temperature of the baths. The ratio of tunneling resistances is fixed to $x=1/3$ and the gate voltage is $V_G=0$.
\label{fig:ALdifT}}
\end{figure}

Note that the only relevant energy scale of the frequency dependence of $S_{\AL}$ is the GL relaxation rate. Fig.~\ref{fig:AL} shows $S_{\AL}$ as a function of the frequency $\epsilon$ for the case $T_1=T_2=1.01 T_c$ and the ratio of tunneling resistances $x=1/3$. The red dashed line presents zero gate voltage and the blue solid line corresponds to $V_G=T_c/2$. One sees that the Aslamazov-Larkin contribution is strongly suppressed by the gate voltage due to suppression of the Cooper pair lifetime.  The influence of temperatures of the thermal baths on the noise is shown in Fig.~\ref{fig:ALdifT}. One observes a similar behavior as in the case of the Maki Thompson contribution.

Finally, in the equilibrium ($T_1=T_2=T$ and $V_G=0$) we find the leading term in the AL ac conductivity using the FDT and Eq.~(\ref{eq:finalAL}) \cite{Schmidt,acMT}
\begin{align}\label{eq:sigmaALI}
\mathrm{Re}\left[\sigma_{\AL}(\epsilon)\right]=&\frac{2e^2}{\pi d_f}\frac{T}{\epsilon}\Bigg[
\arctan{\left(\frac{\epsilon}{2\tau_{\GL}^{-1}}\right)}-
\frac{\tau_{\GL}^{-1}}{\epsilon}\ln{\left(1+\frac{\epsilon^2}{4\tau_{\GL}^{-2}}\right)}
\Bigg].\end{align}
Here the GL rate takes its equilibrium value $\tau_{\GL}^{-1}=8 T\ln{(T/T_c)}/\pi$. We point out that subleading terms to Eq.~(\ref{eq:sigmaALI}) in the dc case logarithmically depend on the GL rate, see Ref.~\onlinecite{ACmy}.

\section{Conclusions\label{sec:conclusions}}

To summarize, we have studied the influence of superconducting fluctuations on the current-current correlation function in a disordered superconducting film driven out-of-equilibrium. The film is assumed to be in the normal state, but close to the nonequilibrium transition into the superconducting state. We obtained and analyzed {nonlinear} dependence of the noise on temperatures of the reservoirs, difference of chemical potentials, and frequency. We first considered the effect of the nonequilibrium conditions on the current-current correlation function (i.e., noise) in the case of noninteracting electrons. Then, we studied the BCS interaction and calculated contributions of different physical nature and origin to the in-plane current fluctuations: the Aslamazov-Larkin given by Eq.~(\ref{eq:finalAL}), the Maki-Thompson given by Eq.~(\ref{eq:MTan}) and the density of states given by Eqs.~(\ref{eq:dosresult}) and (\ref{eq:dosdifT}). These results are new and relevant for future experiments. We find that the frequency dependence of different contributions is characterized by different relevant energy scales manifesting different underlying physical processes. As a special case of our results, in the equilibrium we obtained the ac longitudinal conductivity.

\section*{Acknowledgments}

We acknowledge Z. Ristivojevic for numerous useful remarks. A.~P.~acknowledges the support from the ANR Grant No. ANR-2011-BS04-012 and ANR Grant No. 09-BLAN-0097-01/2.
Work of V.~M.~V.~is supported by the U.S. Department of Energy, Office of Basic Energy Sciences under contract no. DE-AC02-06CH11357.
\appendix

\section{Model\label{app:model}}

In this section we introduce the model that is the starting point for the derivation of the low-energy theory, i.e.~the nonlinear sigma model that is given in Sec.~\ref{sec:NLSM} and used in the rest of the paper. The Hamiltonian can be written as a sum of two parts, $H=H_0+H_{\rm int}$. The single-particle Hamiltonian $ H_0$ reads as
\begin{align}
H_0=\int \dif {\mathbf r}\;\bar\psi_\alpha\left[-\frac{(\bnabla - ie\mathbf{A})^2}{2m} +U_{\rm dis} +e \phi \right]\psi_\alpha,
\end{align}
in the coherent state basis. The fields $\mathbf A$, $\phi$ and $U_{\rm dis}$ are the vector, scalar and disorder potentials, respectively. The electron charge is denoted by $e$, while  $\alpha\equiv\uparrow$,$\downarrow$ stands for spin variable. The summation over the spin indices $\alpha$ is implicitly assumed.
The disorder potential is generated by quenched impurities and it is short-ranged. We assume that it is Gaussian distributed with the variance
\begin{align}
\langle U_{\mathrm{dis}}(\br)U_{\mathrm{dis}}(\br')\rangle =\frac{1}{2\pi\nu\tau}\delta(\br-\br').
\end{align}
Here $\nu$ denotes the bare single particle density of states at the Fermi level per one spin projection and $\tau$ is the elastic scattering time. Also, the disorder is assumed to be weak, i.e. $1/\tau\ll E_F$ where $E_F$ denotes the Fermi energy.
The interaction is given by the Bardeen-Cooper-Schrieffer Hamiltonian
\begin{align}
H_{\rm int}=
-\frac{\lambda}{\nu} \int\dif {\mathbf{r}} \; \bar\psi_{\uparrow}\bar\psi_\downarrow\psi_\downarrow \psi_\uparrow,
\end{align}
where the coupling constant $\lambda$ is positive.

Next, we shortly describe the procedure while the detailed derivation of the nonlinear sigma model is given in Refs.~\cite{Feigelman+00,KamenevAndrev}. One first performs the disorder average. This generates a four-fermion term. Then, one carries the standard decoupling of the four-fermion terms both in the single-particle and the interaction part of the Hamiltonian, via the Hubbard-Stratonovich fields $Q$ and $\Delta$, respectively. Next, the goal is to find a stationary saddle point solution for field $Q$ and to examine the massless fluctuations around it. The massive fluctuations can be integrated out in the Gaussian approximation leading to renormalization of the parameters of the model.  Then, performing the expansion in gradients of $Q$ around the saddle point solution, one obtains the low energy theory valid for energies smaller than the elastic scattering rate and given by Eq.~(\ref{eq:S}).

\section{Density of states contribution to the noise\label{app:B}}

In this section we discuss the density of states contribution to the noise for the case $T_1\neq T_2$, close to the transition into the superconducting state. In Sec.~\ref{sec:DOS} we calculated it for $T_1=T_2$, see Eq.~(\ref{eq:dosresult}). We start from  Eq.~(\ref{eq:dosfinal}) and obtain
\begin{align}\label{eq:dosdifT}
S_{\DOS}(\epsilon)\approx&-\frac{e^2}{\pi^3d_f}
\left[1-\tilde{F}^h(0)\tilde{F}^e(0)\right] T_e^2 z_{cp} \ln{\left(\frac{T_e}{\tau_{\GL}^{-1}}\right)} \Big[
x^2D_1\left(eV_G(1-x),eV_G(1-x),T_1,T_1\right)\notag\\&
+x(1-x)D_1\left(eV_G(1-x),-eV_Gx,T_1,T_2\right)
+x(1-x)D_1\left(-eV_Gx,eV_G(1-x),T_2,T_1\right)\notag\\&
+(1-x)^2D_1\left(-eV_Gx,-eV_Gx,T_2,T_2\right)
+\epsilon\to-\epsilon\Big].
\end{align}
Here the function $D_1$ is defined as
\begin{widetext}
\begin{align}\label{eq:A2}
D_1(x,y,T_1,T_2)=&4\mathrm{Im}\Bigg[\frac{1}{T_1}
\sum_{n\geq0}
\frac{
\tanh{\left(\frac{y-x+\epsilon}{2T_2}+i\frac{2n+1}{2}\pi\frac{T_1}{T_2}\right)}
}{(2n+1-i\frac{-x+\epsilon}{\pi T_1})^2}
+\frac{1}{T_2}
\sum_{n\geq0}
\frac{
\tanh{\left(\frac{x-y-\epsilon}{2T_1}+i\frac{2n+1}{2}\pi\frac{T_2}{T_1}\right)}
}{(2n+1+i\frac{y}{\pi T_2})^2}
\notag\\&-\frac{1}{2T_1}\sum_{n\geq0}\frac{\tanh{\left(i\pi\frac{2n+1}{2}\frac{T_1}{T_2}+
\frac{x+y+\epsilon}{2T_2}\right)}}{i(2n+1)+\frac{x+\epsilon}{\pi T_1}}\frac{1}{i(2n+1)+\frac{x}{\pi T_1}}\notag\\&-\frac{1}{2T_2}\sum_{n\geq0}
\frac{\tanh{\left(i\pi\frac{2n+1}{2}\frac{T_2}{T_1}-
\frac{x+y+\epsilon}{2T_1}\right)}}{i(2n+1)-\frac{y}{\pi T_2}}\frac{1}{i(2n+1)-\frac{y+\epsilon}{\pi T_2}}
\Bigg].
\end{align}
For the validity of this expression see the discussion below Eq.~(\ref{eq:epsilono}). The definitions of the parameters are given in Sec.~\ref{sec:fluctuations1}. The expression \ref{eq:dosdifT} can be easily numerically evaluated. Fig. \ref{fig:dosdifT} shows the influence of temperatures of the thermal baths on the density of states contribution to the noise. One observes a similar tendency as in the case of two other contributions, the Maki-Thompson and the Aslamazov-Larkin, see Sec.~\ref{sec:MT}.

\begin{figure}\includegraphics[width=0.65\columnwidth]{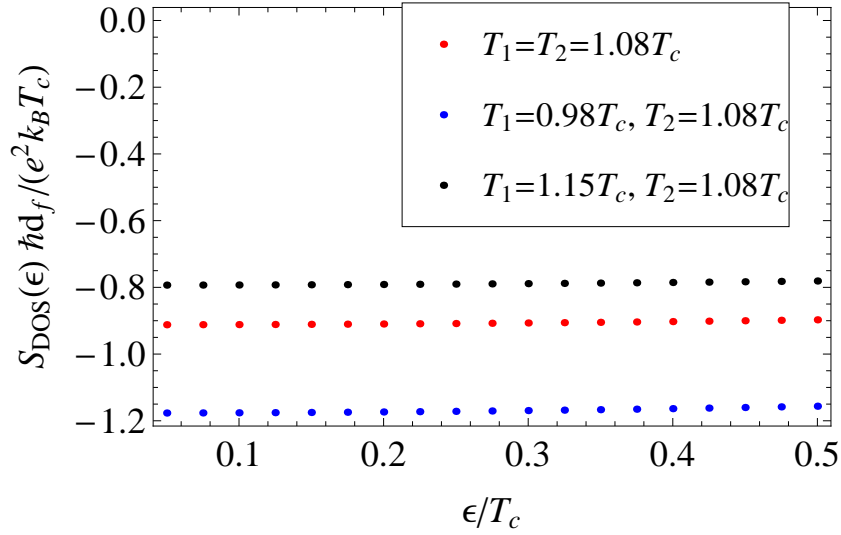}\\
\caption{Current-current correlation function $S_{\DOS}$ as a function of frequency $\epsilon$ for different temperatures of the baths. The ratio of tunneling resistances is fixed to $x=1/3$ and the gate voltage is $V_G=0$.
\label{fig:dosdifT}}
\end{figure}

From Eq.~(\ref{eq:A2}), we obtain the closed-form result for large frequencies, i.e.~in the limit $|\epsilon|,|\epsilon\pm eV_G|\gg T_1,T_2$
\begin{align}
D_1(x,y,T_1,T_2)=&\mathrm{sgn}(y-x+\epsilon)\mathrm{Im}\left[
\frac{1}{T_1}\Psi'\left(\frac{1}{2}-i\frac{-x+\epsilon}{2\pi T_1}\right)
-\frac{1}{T_2}
\Psi'\left(\frac{1}{2}+i\frac{y}{2\pi T_2}\right)\right]
\notag\\&-\frac{\pi}{\epsilon}\mathrm{sgn}{\left({x+y+\epsilon}\right)}
\mathrm{Re}\Bigg[
\Psi\left(\frac{1}{2}-i\frac{x}{2\pi T_1}\right)-\Psi\left(\frac{1}{2}-i\frac{x+\epsilon}{2\pi T_1}\right)
\notag\\&+\Psi\left(\frac{1}{2}+i\frac{y}{2\pi T_2}\right)
-\Psi\left(\frac{1}{2}+i\frac{y+\epsilon}{2\pi T_2}\right)
\Bigg]+\mathcal{O}(e^{-|y-x+\epsilon|/T_i},e^{-|y+x+\epsilon|/T_i}),
\end{align}
where $i=1,2$.
In the case $T_1=T_2=T$, the function $D_1(x,y,T,T)$ becomes equal to $D(x,y)$ defined by Eq.~(\ref{eq:eqD}).
\end{widetext}

%

\end{document}